\documentclass[twocolumn]{pasj01}

\Received{$\langle$reception date$\rangle$}
\Accepted{$\langle$acception date$\rangle$}
\Published{$\langle$publication date$\rangle$}
\draft
\usepackage{soul}
\usepackage{color}

\usepackage{lineno}

\begin{document} 

\Received{2021/12/01}
\Accepted{2022/02/26}

\title{The symbiotic and bipolar nebula M\,2-9: Morphological variability of the collimated ionized wind arising from the core\thanks{\textbf{Accepted for publication in Publication of the Astronomical Society of Japan (2022).} Thesis submitted by Garc\'ia--Flores as a partial fulfillment for the requirements of Bs. Sc Degree in Physics, Licenciatura en F\'isica, CUCEI, Universidad de Guadalajara}}


\author{Eduardo \textsc{de la Fuente}\altaffilmark{1,2,3}
\thanks{Corresponding authors: eduardo.delafuente@academicos.udg.mx and trinidad@astro.ugto.mx}}
\altaffiltext{1}{Departamento de F\'{i}sica, CUCEI, Universidad de Guadalajara, Blvd. Marcelino Garc\'{i}a Barragan 1420, Ol\'{i}mpica, 44430, Guadalajara, Jalisco, M\'exico}
\email{eduardo.delafuente@academicos.udg.mx}
\altaffiltext{2}{Institute for Cosmic Ray Research (ICRR), University of Tokyo, 1--5 Kashiwanoha 5-Chome, Kashiwa, Chiba 277-8582, Japan}
\altaffiltext{3}{Maestr\'ia en Ciencias de Datos, CUCEA, Universidad de Guadalajara, Perif\'erico Norte 799 N\'ucleo Universitario, Los Belenes, 45100 Zapopan, Jalisco, M\'exico}

\author{Miguel A. \textsc{Trinidad}\altaffilmark{4,$\dagger$}}
\altaffiltext{4}{Departamento de Astronom\'{i}a, Universidad de Guanajuato, Apartado Postal 144, 36000, Guanajuato, Guanajuato, M\'exico}
\email{trinidad@astro.ugto.mx}

\author{Daniel \textsc{Tafoya}\altaffilmark{5}}
\altaffiltext{5}{Department of Space, Earth, and Enviroment, Chalmers University of Technology, Onsala Space Observatory, 439 92 Onsala, Sweden}
\email{daniel.tafoya@chalmers.se}

\author{Ivan \textsc{Toledano-Ju\'arez}\altaffilmark{3,6}}
\altaffiltext{6}{Doctorado en Ciencias F\'{i}sicas, CUCEI, Universidad de Guadalajara, Blvd. Marcelino Garc\'{i}a Barragan 1420, Ol\'{i}mpica, 44430, Guadalajara, Jalisco, M\'exico}
\email{toledano.ivan16@gmail.com}

\author{Samuel \textsc{Garc\'ia-Flores}\altaffilmark{7,*}}
\altaffiltext{7}{Licenciatura en F\'{i}sica, CUCEI, Universidad de Guadalajara, Blvd. Marcelino Garc\'{i}a Barragan 1420, Ol\'{i}mpica, 44430, Guadalajara, Jalisco, M\'exico}
\email{samuelo.2@hotmail.com}


\maketitle

\begin{abstract}

We studied the central region of bipolar nebula M\,2-9 using radio-continuum observations obtained from the Jansky Very Large Array (JVLA) and the Atacama Large Millimeter Array (ALMA) interferometers. This work presents new images at $\sim$ 43 GHz ($\sim$ 7.0 mm) and $\sim$ 345 GHz ($\sim$ 0.9 mm) with angular resolutions of $\sim$ 0$\rlap{.}^{\prime\prime}$047 and 0$\rlap{.}^{\prime\prime}$09, respectively. The continuum emission at $\sim$ 43 GHz shows an elongated jet-like structure perpendicular to the $\sim$ 345 GHz observation. We conclude that both emissions could correspond to tracing an isothermal collimated fast wind with constant expansion velocity and being driven by the circumstellar ring traced by ALMA. Although this configuration has been discussed within the scope of planetary nebulae models, there is a remarkable fact: the collimated fast wind shows morphological spatial variability. This supports the idea of a symbiotic binary system within the innermost part of M\,2-9, which would be composed of a white dwarf and an AGB star. The latter could explain the mirror symmetry observed at larger scales due to their orbital motion.

\end{abstract}

\section{Introduction}
\label{sec:intro}

Mass-loss processes happen during the last stages of evolution in low and intermediate-mass stars (M $<$ 8 M$_{\odot}$). They are of significant relevance because they determine the ultimate destiny of these stars and the  potential planetary systems around them. Furthermore, these stars are the progenitors of the vast majority of evolved stars in our Galaxy. Therefore, understanding the mass and chemical composition of this material, that replenishes the interstellar medium, is essential to advance our knowledge of stellar evolution and the chemical evolution of galaxies (e.g. \cite{Tayler1990}).

It is well-known that intermediate and low mass stars end their lives by undergoing a \textit{superwind} in the asymptotic giant branch (AGB) phase (up to several $10^{-5}$ M$_{\odot}$ yr$^{-1}$ where 50 to 80 \% of their stellar mass is lost. The mass-loss rate reaches values as high as 10$^{-4}$ M$_{\odot}$ yr$^{-1}$, which forms a dense, slowly expanding circumstellar envelope (CSE) of dust and molecular gas. Eventually, this mass loss determines the evolutionary pathway of the AGB star. When the star's surface temperature becomes high enough to emit a significant flux of ultraviolet radiation in the white dwarf (WD) stage, it begins to photo-ionize the circumstellar material and forms an optically visible planetary nebula (PN, plural PNe). 

\citet{Kwok1978} propose the first model to explain the formation of PNe using the assumption that the mass-loss from AGB stars is spherically symmetric. This assumption was supported by observations of the distribution of masers and the profiles of molecular lines (see \cite{Kwok2000} and references therein). Nevertheless, PNe are rarely spherically symmetric; typical shapes such as bipolar, butterfly and point symmetric structures are observed.  There are even more complex morphologies such as multipolar configurations and jets (e.g. \cite{Tafoya2020} and references therein). For example, to explain bipolar PNe from outflow collimation by common envelope evolution, \citet{Yangyuxin2020} evokes the presence of fast collimated outflows that carve cavities within the spherical CSE. These cavities are eventually seen as bright lobes and bubble-like morphologies in the subsequent PN phase once the star become hot enough to ionize the CSE (e.g. \cite{Yangyuxin2020} and references therein). 

\begin{figure}
\begin{center}
\includegraphics[width=\columnwidth]{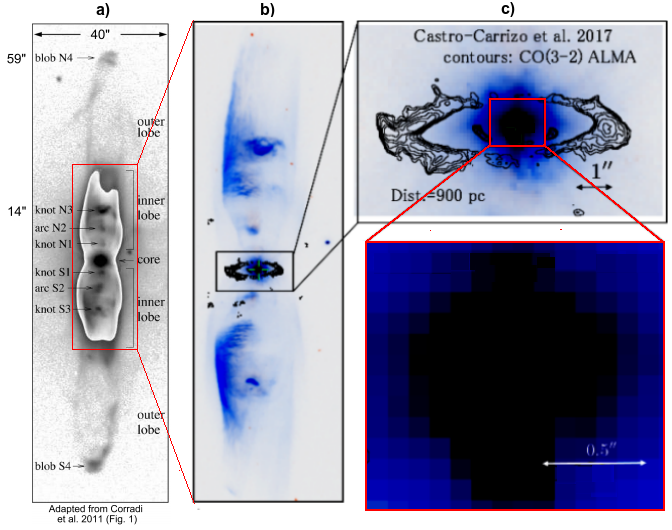}
\end{center}
\caption{\textbf{a)} HST image of the M\,2-9 nebula at outer scales. \textbf{b)} The inset in a) shows the inner structure and core of M\,2-9. These images are adapted from \citet{Corradi2011}. The green cross at b) marks the center of the core. \textbf{c)} The inset in b) shows the core of M\,2-9, with the external molecular disk reported by \citet{Castro-Carrizo2017} in black contours (see their Fig. 3). The red square as inset shows the central part of M\,2-9.}
\label{fig:M2_9}
\end{figure}

\begin{figure*}
\begin{center}
\includegraphics[width=\textwidth]{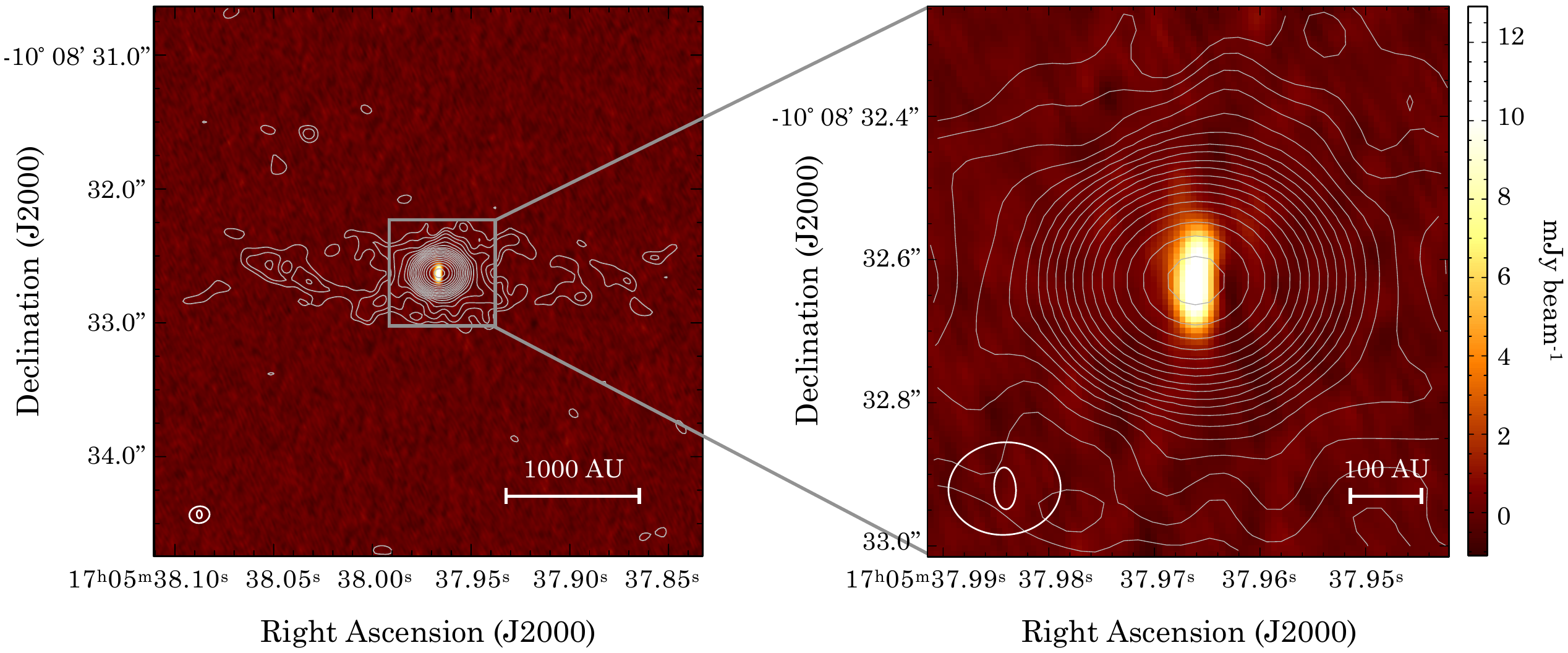}
\end{center}
\caption{The JVLA radio-continuum observation at $\sim$ 43 GHz (colors) for set II (2006) of the innermost part of M\,2-9 (showed as inset) overlapped with the continuum emission of ALMA at 345 GHz (contours) taken in 2013. Contours are drawn from 5-$\sigma$ in steps of 5-$\sigma$ (where $\sigma$ is the rms noise level of 35~$\mu$Jy~beam$^{-1}$). The integrated fluxes for $\sim$ 43 GHz and 345 GHz are 47 mJy and 300 mJy respectively. The beams were 0.055''$\times$0.03477; PA = 6.49$^{\circ}$, and 0.158''$\times$0.129''; PA = -89.80$^{\circ}$ for the JVLA and ALMA observations respectively are shown as ellipses at the bottom left. The flux of ALMA was obtained from inside an ellipse with a major axis of 42$\pm$1 mas, minor axis = 34$\pm$2 mas and PA = 175$\pm$9 degrees after a single Gaussian deconvolution. See Table \ref{tab:tab1}. 
}
\label{fig:alma_vla}
\end{figure*}

\begin{figure}
\begin{center}
\includegraphics[width=\columnwidth]{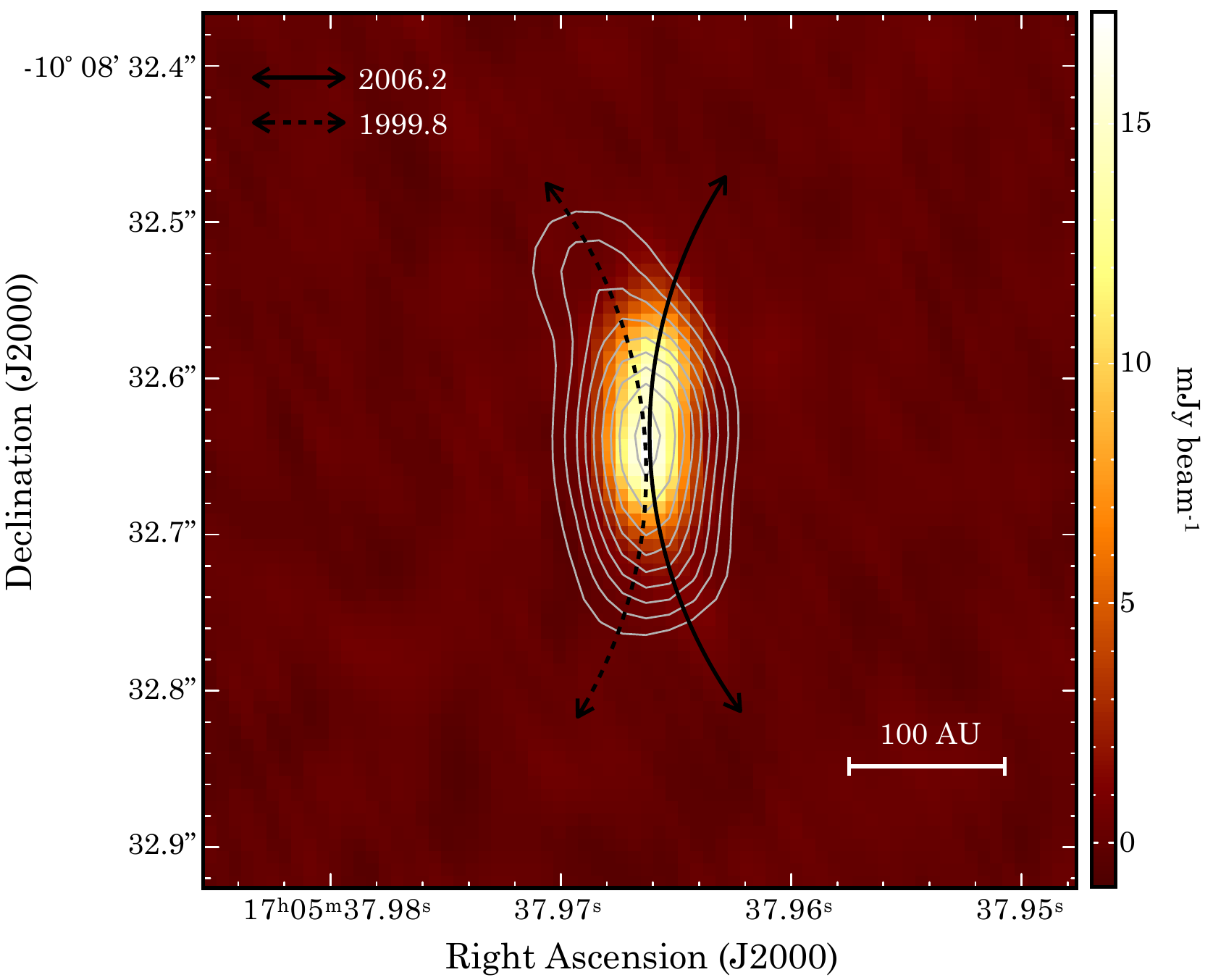}
\end{center}
\caption{The JVLA $\sim$ 43 GHz radio continuum observation shown in Fig. \ref{fig:alma_vla} (color) with an overlay of the $\sim$ 23 GHz emission in white contours (set I; 1999 and reported by \citet{Lim2003}. The black lines show that the direction towards the jet is pointing outwards (dotted for 1999 and solid for 2006). We observe a clear precession of the jet. Contours are from 10\% to 90\% of the peak value (10 mJy beam$^{-1}$). The beam size for both observations, corresponding to the $\sim$ 23 GHz emission of \citet{Lim2003} is 0.108''$\times$ 0.074"; PA = 25.170$^{\circ}$.}
\label{fig:vla_epoch}
\end{figure}

\begin{longtable}{cllcccccc}
\caption{Spectral Energy Distribution Results}
\label{tab:tab1}
\hline 
Frequency  & Flux & Flux\_err & major\_axis & major\_axis\_err & minor\_axis & minor\_axis\_err & PA & PA\_err  \\
 (GHz)  & (Jy) & (Jy) & (mas) & (mas) & (mas) & (mas) & ($^{\circ}$) & ($^{\circ}$) \\ 
\endfirsthead
\hline
\endhead
\hline
\endfoot
\hline
\endlastfoot
\hline
43.34000  & 0.04707 & 0.00376 & 99.1 & 4.5 & 27.8 & 1.5 & 176.7 & 1.1 \\
22.45995  & 0.02481 & 0.00284 & 159.4 & 11.4 & 46.6 & 7.5 & 2.5 & 2.8 \\
14.93962  & 0.01729 & 0.00247 & 234.0 & 8.5 & 40.5 & 12.1 & 178.3 & 1.5 \\
8.459633  & 0.01006 & 0.00147 & 528.3 & 10.7 & 256.2 &  2.9 & 7.7 & 0.59 \\
4.860100  & 0.00787 & 0.00093 & 582.0 & 25.0 & 125.0 & 39.0 & 178.9 & 2.2 \\
\end{longtable}

\begin{figure*}
\begin{center}
\includegraphics[width=\textwidth]{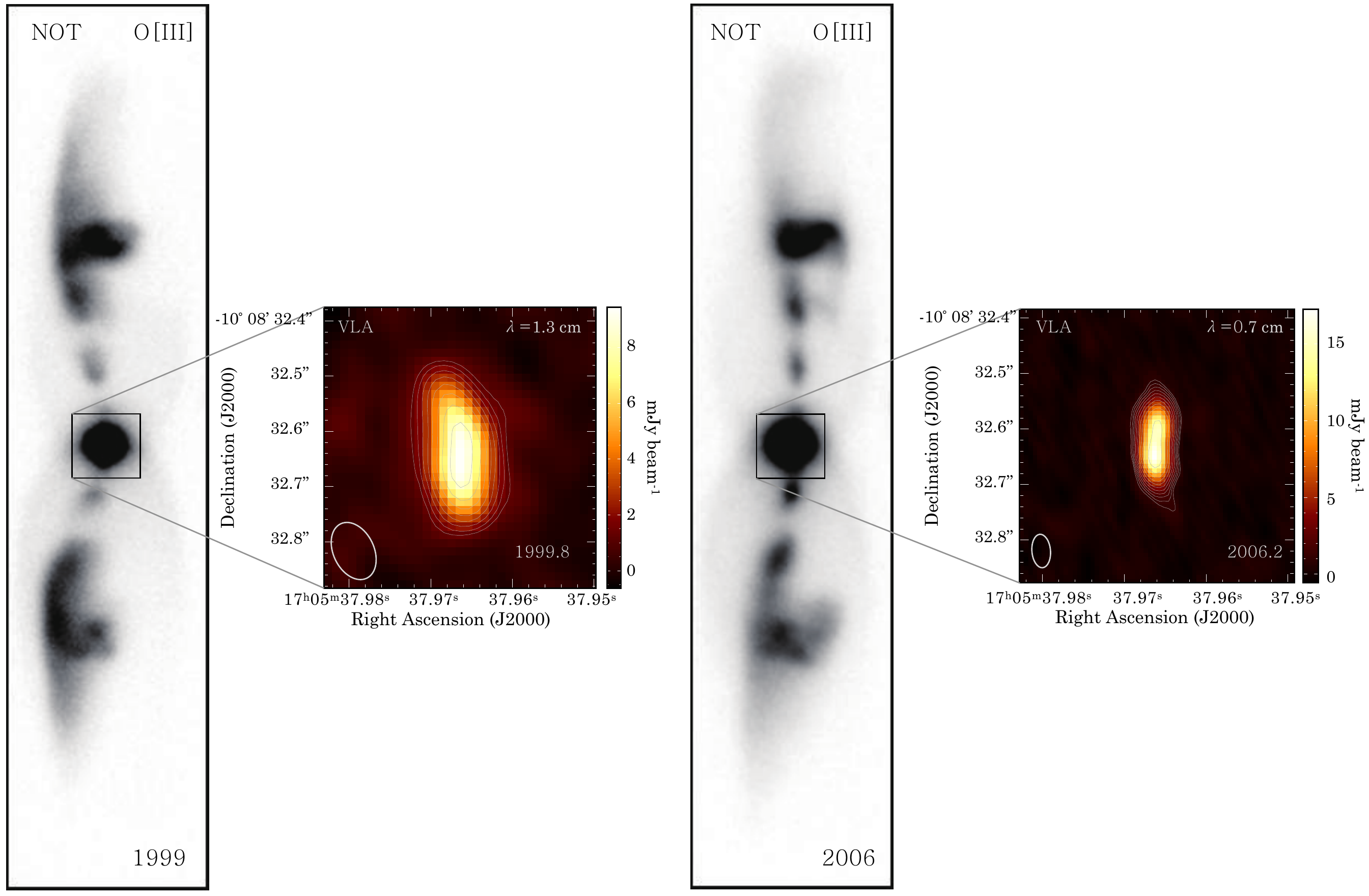}
\end{center}
\caption{\textbf{Left:} Nordic Optical Telescope (NOT) image of M\,2-9 taken from \citet{Corradi2011} in the year 1999, with the $\sim$ 23 GHz emission JVLA contour map from \citet{Lim2003} as inset. Contours are from 10\% to 90\% of the peak value (10 mJy beam$^{-1}$). The beam size of 0.108''$\times$ 0.074"; PA = 25.170$^{\circ}$ is shown in the bottom left. Right: The same as for left, but for 2006 using the JVLA contour map at 43 GHz (set II; 2006) as the inset. Contours are from 10\% to 90\% of the peak value (17 mJy beam$^{-1}$). The true beam size (not the same as in Fig.~\ref{fig:vla_epoch}) of 0.055''$\times$0.03477; PA = 6.49$^{\circ}$ is shown in the bottom left. Both RC emissions show the jet located in the core of M\,2-9; the jet appears to be aligned with the bright regions mentioned in \citet{Livio2001} including knot N1 and arc N2 (compare with Fig.~\ref{fig:M2_9}a). This alignment suggests that the jet is the precursor and driver of the mirror observed at the N3 source (refer to Fig. \ref{fig:M2_9}a).
}
\label{fig:corradi_alma_vla}
\end{figure*}

\begin{figure*}
\begin{center}
\includegraphics[width=\textwidth]{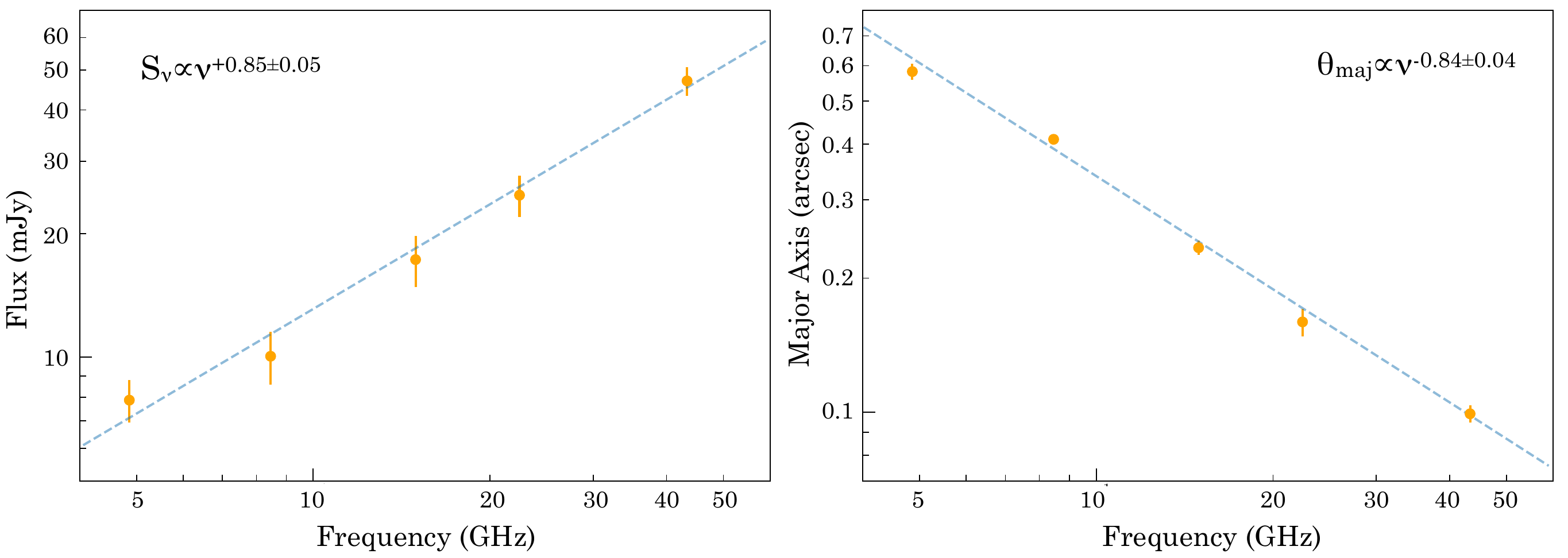}
\end{center}
\caption{Flux density (Left) and major axis size (Right) vs. frequency in log-log of the central core of M\,2-9. The data with respective error bars are shown in yellow. The dotted blue line corresponds to the best fit obtained by using a least squares fit. By adding the 43 GHz point, we can confirm the RC emission as a jet. See text for details. The rest of the points are from \citet{Lim2003}.}
\label{fig:sed}
\end{figure*}

\begin{figure}
\begin{center}
\includegraphics[width=\columnwidth]{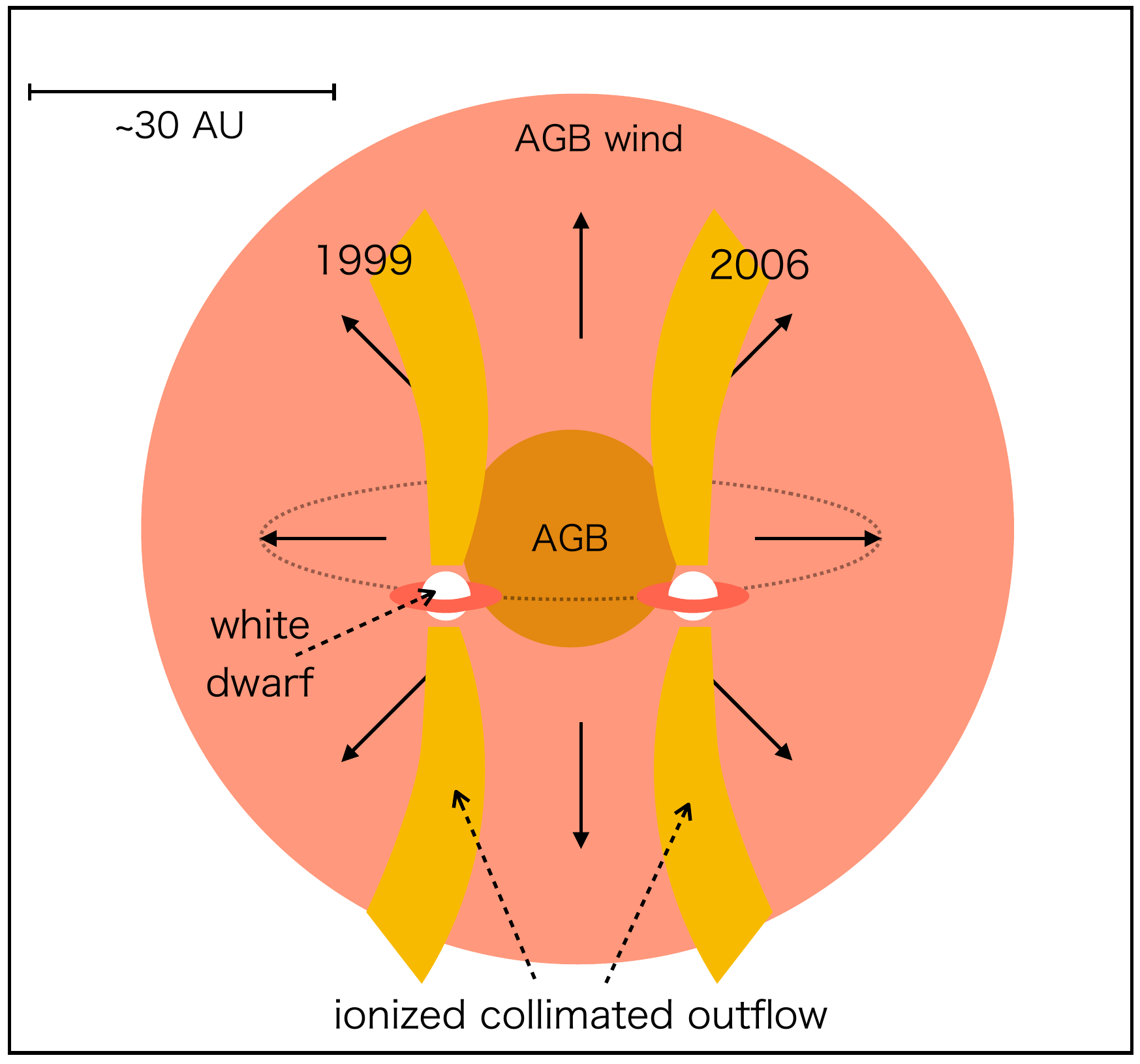}
\end{center}
\caption{Schematic of the \citet{Livio2001} model to explain the \textit{mirror symmetry} discussed by \citet{Corradi2011}. Compare with the RC observations of Figs.~\ref{fig:vla_epoch} and \ref{fig:corradi_alma_vla}, where the ionized collimated outflow is associated to the jet traced at $\sim$ 43 GHz. 
}
\label{fig:livio_alma_vla}
\end{figure}

The \textit{Butterfly Nebula}, \textit{Twin Jet Nebula} (Minkowski 2-9 or M\,2-9; NGC 6302; Caldwell 69; IRAS 17028-1004), is a thoroughgoing object located at $\alpha$(J2000) = 17$^h$ 05$^m$ 37.952$^s$, $\delta$(J2000) = --10$^{\circ}$ 08' 34.58'' within the Scorpio constellation at a distance between 0.81 and 1.17 kpc \citep{Lago2019,Meaburn2008}. It was discovered by William Herschel in 1826\footnote{https://www.nasa.gov/feature/goddard/caldwell-69} and catalogued for the first time by \citet{Minkowski1947}. It is typically classified as a planetary nebula with bipolar morphology (e.g. \cite{Kwok1985}; Fig.~\ref{fig:M2_9}a), although it has also been classified as a symbiotic nebula (e.g. \cite{Ilkiewicz2017}). Therefore, the combination between a symbiotic and bipolar morphologies cannot be discarded (e.g. \cite{Corradi2011,Livio2001}). This object stands out because of two inner lobes at a projected distance of 20'' from the center (e.g. \cite{Livio2001}; see Fig.~\ref{fig:M2_9}a,b), and two fainter outer lobes which can be observed at scales of a few arc-min from the center of the system (e.g. \cite{Schwartz1997}; see Fig.~\ref{fig:M2_9}a). 

Besides, a $\sim$ 7'' in diameter molecular ring (see inset of Fig.~\ref{fig:M2_9}b) surrounding the central object, and both inner and outer bipolar lobes were detected by \citet{Castro-Carrizo2017}. More relevant information of M\,2-9 are presented by \citet{Kastner2021,Balick2018,Livio2001,Lim2000,Doyle2000,Trammell1995,Balick1989} and references therein.

The central part is shown in Fig.~\ref{fig:M2_9}c (inset). \citet{Lim2000} strongly suggest the presence of a WD companion with a mass-loss rate from 10$^{-5}$ to 10$^{-6}$ M$_{\odot}$ yr$^{-1}$ as an ionizing source \citep{Balick1989}. It has an H$\alpha$ line width of 11,000 km s$^{-1}$, which indicate the presence of wind with a velocity of 5500 km s$^{-1}$. \citet{Torres1998} suggests this wind is a result of the presence of an accretion disk around the WD.

Comparing [OIII]($\lambda$5007 \AA) and H$\alpha$+[NII] Nordic Optical Telescope (NOT) observations, \citet{Corradi2011} found an intriguing rotating pattern in the inner lobes resembling a mirror symmetry (see their Fig. 5 comparing panels 1999 and 2008). This behavior has not been clearly explained, although it is widely believed that an involved binary symbiotic-like system with a period of $\sim$ 86 years could explain the phenomena by introducing a rotating and collimated jet launched by the binary system \citep{Corradi2011}. The latter is also supported by \citet{Lykou2011} who confirmed the presence of an inner accretion disk (size $\sim$ 30$\times$40 mas) around the binary system, which was entirely independent from the disk described by \citet{Castro-Carrizo2017}. In addition to this, \citet{Livio2001} modelled this binary system as an AGB or post-AGB start with a WD companion separated by 27 UA, which could explain this mirror symmetry by the motion of this system in an orbital period of about 120 years \citep{Doyle2000}. Using radio-continuum (RC) observations at 1.3, 2, 3.6, and 6 cm obtained with the Karl Jansky Very Large Array (JVLA) in A-configuration, \citet{Lim2003} report the presence of an arc-sec size jet. 

We studied the inner part of M\,2-9 at arc-sec scales by using new continuum observations from the JVLA and Atacama Large Millimeter Array (ALMA) radio interferometers at $\sim$ 43 GHz ($\sim$0.7 cm) and $\sim$345 GHz ($\sim$0.9 mm), respectively. The aim was to investigate the presence of this collimated fast wind suggested by \citet{Livio2001}, with the JVLA, in a disk-like structure modelled by \citet{Lykou2011} and traced by ALMA. The latter allowed us to identify the nature of this nebula and its location with the correct evolutionary schema. This study also includes a spectral energy distribution (SED) analysis to complement the work of \citet{Lim2003} and determine the nature of RC observations. Observations are described in Section~\ref{sec:obs}. Results and discussion are shown in Section~\ref{sec:res_disc}. Conclusions are summarized in section~\ref{sec:conclusion}.


\section{JVLA and ALMA Observations}
\label{sec:obs}

We retrieved RC observations of M\,2-9 from the NRAO-JVLA data archive for two different sets: $\sim$ 23 GHz (1.3 cm) for set I (the year 1999; published in \cite{Lim2003}) and $\sim$ 43 GHz for the set II (the year 2006; this work). 

The observations for set I (project code: AL0496) were carried out on 1999 September 21st using the JVLA in A-configuration. All the RC observations in this project (1.3, 2, 3.5, and 6 cm) were published by \citet{Lim2003}. All emissions traced to the same region and presented the same elongated morphology in N--S direction, slightly blending towards the NE and SE (see their Fig. 1).

For set II (project code: AL0675), the observations were carried out in 2006 February 14th using 27 antennas of the JVLA in A-configuration, which has a maximum and minimum baseline of 36.4 km and 0.68 km, respectively. The corresponding angular resolution and largest angular scale at the frequency of observation are 0$\rlap{.}^{\prime\prime}$043 and 1$\rlap{.}^{\prime\prime}$2, respectively. The total on-source observation time was 50 minutes. These observations were performed in full polarization mode with two 50 MHz-wide spectral windows. The central frequencies from the spectral windows were 4885.1 MHz and 4835.1 MHz, respectively. The absolute flux calibrator was 1331+305 (with an assumed flux of 1.53 Jy), and the phase calibrator was 17330-13048 (bootstrapped flux of 2.86 Jy). The size of the synthesized beam, using a robust weighting parameter of 0, was 60$\times$33~mas with a P.A. of 5.8$^{\circ}$. The rms noise of the final continuum image is 0.1 mJy beam$^{-1}$.

Calibration and data reduction of the JVLA observations were performed by following standard procedures in AIPS. After the phase-reference calibration, self-calibration in phase was also performed.

In addition, we obtained archival RC ALMA data (project code: 2013.1.00458.S, PI: A. Castro-Carrizo) at $\sim$ 345 GHz, to study the emission produced by the dust associated with M\,2.9. The observations were carried out in three separate sessions on July 4, 10 and 11, 2013, with 42, 46 and 45 antennas, respectively. The ALMA 12 m array with band 7 receivers ($\sim$350~GHz) were used. Thus, the minimum and maximum baseline lengths were 15.3 m and 3.6 km, which provided a nominal angular resolution and maximum recoverable scale of 0$\rlap{.}^{\prime\prime}$09 and 1$\rlap{.}^{\prime\prime}$2, respectively. The field of view is $\sim$18$^{\prime\prime}$, and the total time on science target was 1 hr and 56 min. The average precipitable water vapor level during the observations was around 0.6 mm. 

We calibrated the data using the ALMA pipeline (version 2020.1.0.40; CASA 6.1.1.15), with J1924$-$2914 as the amplitude and bandpass calibrator (flux density=2.4~Jy at 355~GHz, and spectral index $\alpha$=$-$0.586). Phase calibration was performed using J1851+0035 (flux density=279~mJy at 355~GHz). Atmospheric variations during observations were corrected by using water vapor radiometer data. The calibrated data set contained 4$\times$1.85 GHz spectral windows with 2000 channels each.

\section{Results and discussion}
\label{sec:res_disc}

In Fig.~\ref{fig:alma_vla} we show the 43 GHz (colors) map for set II, overlapped with the ALMA continuum image at 345 GHz (contours). The field of view is $\sim$ 4 arcsec$^2$. The elongated external contours correspond to the dust ring discussed by \cite{Castro-Carrizo2017} (see their Fig. 8). This dust ring coincides with the inner molecular ring observed, whereas the disk-like structure in Fig.~\ref{fig:M2_9}b,c corresponds to the outer molecular ring (see Fig. 3 of \cite{Castro-Carrizo2017}). In the inset we show the central part of M\,2-9, where the elongated structure (north-south direction) from the JVLA image is observed perpendicular to the contours overlapped by ALMA (tracing a more inner dust ring as is reported by \cite{Castro-Carrizo2017}).

We present the $\sim$ 43 GHz image in Fig.~\ref{fig:vla_epoch}, with an overlay of the $\sim$ 23 GHz RC emission adapted from \citet{Lim2003}. Both RC emissions have the same morphology in the central part (elongation at the north-south direction), but the external emissions change in direction, presenting a morphological variability; the $\sim$ 23 GHz emission blends to the left side or direction, while the $\sim$ 43 GHz does so towards the right. The respective black curved arrows in Fig.~\ref{fig:vla_epoch} indicate the latter. Similar behaviour is also observed on the N3, and S3 knots of the HST images (Fig. \ref{fig:M2_9}a); the mirror symmetry discussed by \cite{Corradi2011} (see Fig.~\ref{fig:corradi_alma_vla}). Therefore, we suggest that the jet traced at $\sim$ 23 GHz and $\sim$ 43 GHz is not only correlated with this mirror symmetry, but is precursor and driver (both blends).

\citet{Lim2003} report flux density (S$_{\nu}$) and size ($\Theta$) plots as function of the frequency at 4.86, 8.46, 14.94, and 22.46 GHz (see Tab.~\ref{tab:tab1}), indicating that S$_{\nu}$ $\propto$ $\nu^{0.67}$ and $\Theta$ $\propto$ $\nu^{<-0.86>}$. These values for the S$_{\nu}$ and $\Theta$ spectral indexes are consistent with the theoretical values proposed by \citet{Reynolds1986} of 0.6 and -0.7 for an isothermal wind with a constant expansion velocity. Thus, to investigate the nature of the 43 GHz RC emission shown on Fig.~\ref{fig:vla_epoch}, we present the SED reported in \citep{Lim2003}, adding the new $\nu$ $\sim$ 43 GHz emission (see Fig.~\ref{fig:sed} and Tab.~\ref{tab:tab1}). We confirm the RC emission from 4.86 to 43.34 GHz as a jet, where the free-free emission is also traced at $\sim$ 43 GHz.

In light of the above,  we evoke the \citet{Livio2001} model to explain the morphological variability of the jet observed in Fig.~\ref{fig:vla_epoch} and the relation of this bending with the mirror symmetry observed at larger scales. In the Fig.~\ref{fig:livio_alma_vla} we show a schematic of the M\,2-9 system based on this model. In the \citet{Livio2001} model (see their Figure 1b), the AGB or post-AGB star is expelling material, and some of it falls towards the WD through an accretion disk that fills the respective Roche lobe. 

The outward bipolar jet (the JVLA $\sim$ 43 GHZ RC emission) is the result of magnetocentrifugal launching produced by the accretion disk around the WD. However, as the AGB star loses material isotropically, the jet interacts with the AGB wind, and by hydrodynamic pressure, the AGB wind bends the jet towards the radial direction. 

Although our observations do not have sufficient angular resolution to determine dynamical parameters like the orbital period, we can say that since the time interval between the two observations is around 7 years and taking into account an orbital period of 86 years \citep{Doyle2000,Corradi2011}, the angle subtended by the white dwarf in those epochs, as seen from the red giant star, is around 30 degrees. Furthermore, since the jet exhibited a change in the direction of its bending during that interval of time, it can be inferred that the white dwarf star moved from the left to the right of the red giant star as seen from our perspective (see Fig.~\ref{fig:livio_alma_vla}).

In agreement with \citet{SC2017}, and comparing this to the SED presented by them for M2-9 (see their Figure 1), the 345 GHz emission could be a dust ring produced by the WD during its AGB phase (AGB wind; see \cite{Ferrarotti2006,Kwok2000,Kwok1993,Kwok1980} and references therein) in a binary system. 

Finally, as the WD and the AGB star are orbiting each other, the jet changes its orientation as the WD is moving relative to the AGB star. Therefore, the jet is bent by the wind from the AGB star, contributing to the mirror symmetry discussed by \citet{Corradi2011}. In this context, the presence of the jet confirmed in this work was the missing piece on the puzzle to be able to apply the \citet{Livio2001} model and test it.

\section{Conclusion}
\label{sec:conclusion}

\begin{itemize}

\item Based on the observed SED from 4.8 to 43.3 GHz, we confirm that the RC emission at 43.34 GHz is consistent with a collimated thermal jet with constant expansion velocity. This jet is produced and driven from an accretion disk associated with the WD and related to the 345 GHz emission.

\item The jet traced at 43.34 GHz is blended in opposite directions from 1999 to 2006 (7 years at least). This behavior is in agreement with the mirror symmetry observed on knots N3 and S3 at inner scales of M\,2-9.

\item A symbiotic system is present in the center of M\,2-9, and the mirror symmetry observed at larger scales ($\sim$ 20 arc-sec) can be explained by the inclusion of the jet at 43.34 GHz in the \citet{Livio2001} model. Thus, the jet is the driving source of this mirror symmetry.

\end{itemize}


\begin{ack}

EdelaF thanks: 1.- Colegio departamental de F\'isica, and authorities of CUCEI and UdeG, for authorizing and supporting his sabbatical stay at the Institute for Cosmic Ray Research (ICRR) of the University of Tokyo (UTokyo) in 2021; 2.- Carlos Iv\'an Moreno, Cynthia Ruano, Rosario Cedano, Diana Naylleli Navarro, Luis Fernando Gonz\'alez Bola\~nos, Jorge Alberto Rodr\'iguez Castro, Johnatan Barcenas Alvarez, Patricia Retamoza Vega, Laura Gonzalez Jaime, and Dulce Ang\'elica Valdivia Ch\'avez for administrative support and help during this stay, and 3.- Prof. Takaaki Kajita, Sugimoto Kumiko, Michiru Ito, Shiraga Ryoko, the administrative staff of ICRR; Takashi Sako, Masato Takita, Kasumaza Kawata, students, and the academic staff of ICRR for the invitation, academics, and support. EdelaF and SG--F thanks PRODEP-SEP through a cuerpo acad\'emico (UDG-CA-499) grant, and Coordinaci\'on General Acad\'emica y de Innovaci\'on (CGAI-UDG) for management and financial support; origin of the associated Physics Bs. Sc. Thesis. IT--J acknowledges support from CONACyT, Mexico; grant 754851. This paper makes use of the following ALMA dataset ADS/JAO.ALMA\#2015.1.01539.S. ALMA is a partnership of ESO (representing its member states), NSF (USA) and NINS (Japan), together with NRC (Canada) and NSC and ASIAA (Taiwan) and KASI (Republic of Korea), in co-operation with the Republic of Chile. The Joint ALMA Observatory is operated by ESO,AUI/NRAO and NAOJ.

\end{ack}

\end{document}